\documentclass[
  10pt,
  authoryear,
  5p
]{elsarticle}

\usepackage{amsmath,amsfonts}
\usepackage{amsthm}
\usepackage{comment} 

\usepackage{graphicx}
\usepackage{booktabs}
\usepackage{threeparttable}
\usepackage{longtable}
\usepackage{booktabs}
\usepackage{siunitx}
\usepackage[font=small,labelfont=bf]{caption}
\usepackage{balance}
\usepackage{subcaption}
\usepackage[inline,shortlabels]{enumitem}
\setlist{nosep}

\usepackage{titlesec}
\titlespacing*{\section}{0pt}{1.2ex}{0.8ex}
\titlespacing*{\subsection}{0pt}{1.0ex}{0.6ex}

\usepackage[dvipsnames]{xcolor}
\PassOptionsToPackage{
  colorlinks=true,
  linkcolor=MidnightBlue,
  citecolor=MidnightBlue,
  urlcolor=MidnightBlue
}{hyperref}
\usepackage{hyperref}

\setcitestyle{aysep={}}

\makeatletter
\def\ps@pprintTitle{%
  \let\@oddhead\@empty
  \let\@evenhead\@empty
  \let\@oddfoot\@empty
  \let\@evenfoot\@oddfoot}
\makeatother
\journal{~}

\titleformat{\section}
  {\large\bfseries}     
  {\thesection}{1em}{}

\titleformat{\subsection}
  {\normalsize\bfseries}
  {\thesubsection}{1em}{}

\titlespacing*{\section}{0pt}{1.2ex}{0.8ex}
\titlespacing*{\subsection}{0pt}{1.0ex}{0.6ex}

\begin{document}

\hypersetup{ 
  colorlinks=true,
  linkcolor=MidnightBlue,
  citecolor=MidnightBlue,
  urlcolor=MidnightBlue
}


\begin{frontmatter}

\title{Limits of Predictability in Civil Litigation}

\author[1,2,3]{Sandro Claudio Lera\corref{cor}}
\author[2]{Shahrokh Firouzi}
\author{Jonathan Habshush}
\author[1,3]{Robert Mahari}

\cortext[cor]{Corresponding author: \texttt{slera@mit.edu}}

\address[1]{CodeX, The Stanford Center of Legal Informatics, Palo Alto, USA}
\address[2]{Institute of Risk Analysis, Prediction and Management, Southern University of Science and Technology, Shenzhen, China}
\address[3]{Connection Science, Massachusetts Institute of Technology, Cambridge, USA}

\begin{abstract}
Lawyers, insurers, and litigation funders routinely forecast how civil lawsuits will end, and legal theory distinguishes easy cases, in which the law determines the result, from hard cases, in which it does not. 
Yet there is little systematic evidence on how accurately the outcome of a lawsuit can be forecast from the court record while the case is still pending. 
This article provides such evidence at scale. 
Using 835{,}190 filings and orders from 102{,}721 civil cases in federal and state courts across 12 U.S.\ states, we estimate at each filing the probability that a case ends in a plaintiff win, a plaintiff loss, or a settlement, using only the record available at that point. 
The forecasts draw on structured information extracted from each filing, on comparable earlier cases and how they were resolved, a comparison that practitioners make from experience and that language models now allow to be made systematically from the full text of the record, and on the case histories of the judges and law firms involved. 
We measure how predictable a case is by how concentrated its forecast is on a single outcome. 
Predictability varies systematically across disputes: tort cases are the most predictable, and state-court cases are more predictable than federal ones, whereas business and corporate, intellectual-property, and property disputes and cases involving corporate parties are the least. 
Across all cases, 61\% of forecasts are correct, rising to 95--97\% among forecasts that give a win or a loss more than 90\% probability. 
Richer information about the facts, the judge, and counsel improves forecasts for predictable cases but adds little for unpredictable ones. 
Once case length is taken into account, forecasts do not become more certain as the record grows, and cases that last longer are harder to predict from the outset. 
Settlement is most frequent among cases of intermediate predictability. 
These results provide a benchmark for how predictable civil litigation is, lay the groundwork for an empirical approach to legal indeterminacy, and indicate where outcome forecasts could narrow information gaps between litigants.
\end{abstract}


\end{frontmatter}

\section{Introduction}

Civil litigation is a central mechanism through which economic and social conflicts are formalized and resolved. 
Every participant in it acts on a forecast. 
A plaintiff deciding whether to file, a defendant weighing an offer of settlement, counsel advising on a dispositive motion, an insurer setting a reserve, and a litigation funder pricing a claim all rely on an assessment of how the case is likely to end. 
\citet{Holmes1897} defined the law itself as a prediction of what courts will do. 
Yet these assessments are formed informally, case by case, and the available evidence suggests that they are often poorly calibrated \citep{GoodmanDelahunty2010}.

Reliable evidence on how accurately the outcome of a lawsuit can be forecast from the court record while the case is still pending matters for legal theory as much as for practice. 
Economic accounts of litigation hold that parties bargain in the shadow of the expected judgment, and that disputes proceed to adjudication when the parties' assessments of the merits diverge \citep{Priest1984}. 
Jurisprudence distinguishes easy cases, in which the governing law settles the result, from hard cases, in which it does not, and has long debated how common each is \citep{Fischman2021}. 
Both literatures turn on a quantity that is rarely observed, namely how uncertain the outcome of a given dispute is at a given point in the proceedings. 
That uncertainty is not fixed. 
Parties revise their views of liability, damages, and litigation risk as pleadings are filed, motions are decided, and the record develops, and most disputes end in settlement or in a disposition short of trial. 
Predictability is therefore tied not only to adjudication, but also to a process in which information is revealed gradually and decisions are made sequentially.

Advances in machine learning and natural language processing have renewed interest in predicting legal outcomes. 
Judicial decisions have been forecast in constitutional, appellate, administrative, and civil settings from case metadata, the text of court documents, and the identities of judges and counsel \citep{Katz2017}. 
More recently, calibrated forecasts have begun to be used as empirical instruments for studying legal institutions, including the guidance function of law and the selection of disputes for litigation \citep{Stiglitz2026}. 

Existing work nonetheless leaves this question largely open, for at least four reasons. 
First, many studies rely on information that becomes available only with the decision, such as the opinion itself or the facts as found by the court, so the prediction does not reflect what a litigant knew when decisions had to be made \citep{Medvedeva2023}. 
Second, samples of adjudicated cases are selected. 
They capture the disputes that proceeded to judgment and omit the large share that settled, and most prediction tasks are accordingly framed as a binary choice between judgment for the plaintiff and judgment for the defendant, which abstracts from settlement as a disposition in its own right. 
Third, empirical applications concentrate on appellate and apex courts, above all the U.S.\ Supreme Court, although the great majority of disputes, and much of their economic weight, arise in ordinary civil litigation. 
This leaves little evidence on the mass of disputes in which litigants, counsel, insurers, and courts make routine but consequential decisions. 
Fourth, the information on which practitioners base their assessments has been hard to measure. 
Lawyers judge a case largely by analogy, comparing its facts and arguments with those of cases they have seen before. 
Empirical work has typically approximated this information with coarse categories, such as the area of law or the type of party. 
Language models now make it possible to represent the full text of each filing and to identify, for every case, the earlier cases that most closely resemble it and how they were resolved, systematically and at scale.

These limitations separate existing predictive models from the position of the practitioner. 
Counsel and litigants decide under incomplete and evolving information, drawing on the complaint and answer, subsequent filings and orders, the procedural posture of the case, the outcomes of similar earlier disputes, and the identities of the parties, opposing counsel, and the court. 
Capturing this temporal structure is necessary for understanding not only whether outcomes are predictable, but also at what stage and for which disputes predictability emerges or breaks down. 
The sources of predictability are also likely to be heterogeneous, reflecting the facts of the case, the area of law, the forum, and the conduct of the parties.

In this study, we measure the predictability of civil litigation outcomes in a way that follows this evolving record. 
We assemble 102{,}721 civil cases and 835{,}190 filings and orders from state and federal courts in 12 U.S.\ states.
Each case is represented by the sequence of documents filed by the parties or issued by the court. 
At each filing, we estimate the probability that the case ends in a plaintiff win, a plaintiff loss, or a settlement, using only what was on the record at that point: the content of the documents, the characteristics of the parties, the law firms and judge involved, and the outcomes of similar previously resolved cases. 
This comparison with similar earlier cases formalizes the reasoning by analogy on which practitioners rely, and it turns out to carry the strongest predictive signal in our data. 
By evaluating these estimates on later cases, we provide, to our knowledge, a first large-scale measurement of how accurately civil-litigation outcomes can be forecast before disposition.

Our interest is less in aggregate predictive accuracy than in the uncertainty that remains in each case. 
For each document, the model produces a probability distribution over the three outcomes. 
We use the entropy of this distribution as a measure of outcome uncertainty and interpret it as a model-based proxy for case complexity. 
We use predictability in two related senses: across cases, the share of outcomes that can be forecast correctly from the record; for an individual case, how concentrated its forecast is on a single outcome. 
An easy case, in this sense, is one whose disposition can be anticipated with confidence from the record, and a hard case is one whose disposition cannot. 
This allows us to characterize how uncertainty varies across areas of law, party configurations, and forums, how it evolves over the course of the proceedings, and how it relates to the likelihood of settlement.

More broadly, our approach connects the prediction of legal outcomes with economic theories of litigation and jurisprudential accounts of legal determinacy. 
By treating adjudication and settlement as jointly embedded in a process of information revelation, we move from predicting who prevails to measuring where, and at what stage, the outcome of a civil dispute is uncertain. 
Because divergent expectations are a main reason why disputes fail to settle, a measure of how predictable a case is also matters for the efficiency of litigation and for how evenly information is shared between parties, insurers, and funders.

\section{Related Literature}
\label{sec:literature}

Our study draws on four bodies of work: the jurisprudential debate over easy and hard cases, the economic analysis of selection and settlement, empirical research on how civil cases are resolved, and the prediction of legal outcomes.

\subsection{Easy cases, hard cases, and legal indeterminacy}

The idea that law can be understood as a prediction of what courts will do goes back to \citet{Holmes1897}. 
How reliable such predictions can be has since been debated as a contrast between easy cases, in which the applicable rules settle the result, and hard cases, in which they do not \citep{Dworkin1975,Schauer1985,Marmor1990}. 
Contributions to the related debate over legal indeterminacy argue that indeterminacy is real but moderate \citep{Solum1987,Kress1989}.

How many cases are easy is an empirical question, and the evidence is thin. 
Judges tend to report that only a small minority of the cases before them are hard \citep{Howard2014,Posner2010}, whereas selection models predict the opposite, since easy cases should settle. 
\citet{Fischman2021} shows that the usual indicators, such as dissent rates, do not identify the share of easy cases in appellate courts. 
More recent work measures difficulty through judges' own ratings \citep{Sheppard2026} or through calibrated forecasts of Supreme Court decisions \citep{Stiglitz2026}. 
All of this evidence comes from appellate courts. 
We ask the same question about ordinary civil litigation, including cases that settle, using only information available before the outcome is known.

\subsection{Selection and settlement under uncertainty}

Economic models of litigation make uncertainty about the outcome the reason why some disputes are tried while others settle \citep{Shavell2004,Posner2007,Spier2007,Lera2026}. 
In the model of \citet{Priest1984}, a dispute is tried when the parties' estimates of the plaintiff's chance of success diverge by more than the savings from settling. 
This is most likely for disputes close to the decision standard, so tried cases are a selected sample of uncertain disputes. 
Other models attribute the failure to settle to private information and study how bargaining unfolds before trial \citep{Bebchuk1984,Spier1992,Bebchuk1996}. 
Later theory qualifies what can be inferred from plaintiff win rates \citep{Shavell1996,Klerman2014,Gelbach2018,Klerman2025}.

Empirical tests rely on what is observed, namely trial rates and plaintiff win rates, and support selection without a uniform fifty-percent win rate \citep{Eisenberg1990,Waldfogel1995,Siegelman1995,Kessler1996}. 
Other studies examine which cases fail to settle \citep{Gross1991,Lederman1999} and how rulings on motions change the timing of settlement \citep{Boyd2013}. 
The parties' own forecasts are unreliable, since litigants and their lawyers tend to be self-serving and overconfident in assessing their cases \citep{Loewenstein1993,Kiser2008,GoodmanDelahunty2010}.

The quantity at the center of these models, the probability of each outcome as it can be assessed at a given point in a case, has had to be inferred from trial rates and win rates. 
We estimate it for each case at each filing from the public record. 
Our estimate is not a party's belief, since the parties hold private information, but it allows the relation between outcome uncertainty and settlement to be examined directly.

\subsection{How civil cases end}

A measure of predictability depends on what counts as an outcome. 
Trials have become rare \citep{Galanter2004}, but the cases that do not reach trial are not all settlements \citep{Galanter1993,Clermont2002}. 
\citet{Eisenberg2009} estimate the settlement rate at about two-thirds in two federal districts, with wide variation across case categories. 
Administrative disposition codes are frequently wrong, and establishing from court records how a case ended is itself difficult \citep{Hadfield2004,Vacek2019}.

State courts, where most civil litigation takes place, are dominated by low-value, routine claims, often brought by high-volume corporate plaintiffs against unrepresented individuals \citep{HannafordAgor2015,WilfTownsend2022}, in line with the advantages of repeat players described by \citet{Galanter1974}. 
Some outcomes may therefore be predictable because of who the parties are and how routine the claim is. 
We accordingly include jurisdiction and party type among the correlates of predictability, and treat settlement as an outcome of its own.

\subsection{Predicting legal outcomes}

Statistical prediction of court decisions has a long history in legal scholarship, most of it concerned with the Supreme Court \citep{Nagel1965,Ruger2004,Katz2017}. 
Prediction has since been described as part of ordinary legal practice, with consequences for civil procedure \citep{Katz2013,Engstrom2021}. 
Closer to our setting are studies of settlement in securities class actions \citep{McShane2012} and of trial court outcomes in a single doctrinal area \citep{Alexander2019}.

Another stream of literature predicts decisions from text, for example for the European Court of Human Rights \citep{Aletras2016,Medvedeva2020}, for motions in civil cases \citep{Mcconnell2021}, and with the help of similar earlier cases \citep{Ashley2009,Cao2024,Valvoda2024}, but not to measure how predictable ordinary civil cases are. 
The identity of judges and the work of counsel carry predictive information of their own \citep{Chen2017,Tippett2021,Mahari2024,Mojon2025}.

This literature has well-known limits.\footnote{Surveys of the computational literature include \citet{Alcantara2022,Zhang2025,Dina2025,Santosh2024,Ariai2025}.} 
Many studies rely on information that becomes available only with the decision, such as the judgment itself or the facts as found by the court \citep{Medvedeva2023,Liu2025,Kapoor2024}. 
Most concern apex or appellate courts, or a single court or area of law. 
And most frame the task as a choice between two adjudicated results, which leaves out settlement.

In most of this work, prediction is the goal. 
\citet{Stiglitz2026} instead treats calibrated forecasts of Supreme Court decisions as a measuring instrument, comparing predictability across areas of law and testing the selection hypothesis. 
We apply this approach to ordinary civil litigation. 
Our model uses only what had been filed at each point in a case, covers a broad range of case types in state and federal courts, and includes settlement among the outcomes. 
The object of interest is the uncertainty that remains: how it varies across disputes, how it changes as a case proceeds, and how it relates to settlement.

\section{Data and Methods}
\label{sec:data}

This section describes the sample and the information recorded for each case, and sets out the prediction task.

\subsection{Data and features}

The unit of analysis is the civil case, which we observe through the documents filed in it: the complaint and the answer, motions and briefs, affidavits and declarations, and the orders and judgments of the court. 
We therefore represent each case as the sequence of its documents in the order of filing, so that at any point in the proceedings its record consists of the documents filed so far.

We randomly sampled 102{,}721 civil cases filed between 1996 and 2022 in state and federal courts across 12 U.S.\ states, which together contain 835{,}190 filings and court orders (Table~\ref{tab:sample}). 
The dockets and documents originate from the federal judiciary's PACER system and from the electronic records of state courts, and were obtained through commercial providers of court records. 
The cases fall into seven areas of law: tort, contract, employment, intellectual property, business and corporate, bankruptcy, and property and real estate. 
Each case ends in one of three outcomes, defined below: plaintiff loss (44\% of cases), plaintiff win (22\%), or settlement (34\%). 

\begin{table}[htb]
\centering
\small
\caption{Sample overview.}
\label{tab:sample}
\begin{tabular*}{\columnwidth}{@{\extracolsep{\fill}}p{0.40\columnwidth}p{0.52\columnwidth}@{}}
\toprule
\multicolumn{2}{@{}l}{\textit{Coverage}} \\
\quad Cases & 102{,}721 \\
\quad Filings and orders & 835{,}190 \\
\quad Period of filing & 1996--2022 \\
\quad U.S.\ states & 12 \\
\quad Jurisdictions & State and federal \\
\midrule
\multicolumn{2}{@{}l}{\textit{Case outcome (share of cases)}} \\
\quad Plaintiff loss & 44\% \\
\quad Plaintiff win & 22\% \\
\quad Settlement & 34\% \\
\midrule
\multicolumn{2}{@{}l}{\textit{Prediction sample}} \\
\quad Training set & 82{,}176 cases; 691{,}746 documents \\
\quad Test set & 20{,}545 cases; 143{,}444 documents \\
\quad Features per document & 98 (see \ref{app:features}) \\
\bottomrule
\end{tabular*}
\begin{flushleft}
\footnotesize \textit{Notes:} Cases are randomly sampled from those filed in the covered states over the period. 
Cases are split chronologically by filing date: the earliest 80\% form the training set and the latest 20\% the test set.
\end{flushleft}
\end{table}

We convert each document from PDF to text and prompt a large language model with a fixed set of questions about it. 
The questions cover, among other things, the area of law, the parties, the relief sought, the evidence referred to, the status of pending motions, and, where the document records it, the final disposition of the case. 
The model also writes a summary of the document of at most 500 characters. 
\ref{app:processing} describes the processing in more detail.

Each document is then described by features that use only what was on the record up to that point. 
The features fall into three classes, and \ref{app:features} gives a complete overview.

The first class records the basic structure of the case and the document. 
It includes the jurisdiction (state or federal court), the area of law, and the plaintiff and defendant types (one individual, several individuals, a corporation, or a government entity). 
It also records the type of document and the party filing it, the procedural posture, such as the status of a motion to dismiss or a motion for summary judgment, case-specific indicators, such as insurance involvement or an admission of liability, and timing, such as the time elapsed since the case began.

The second class captures the facts and legal arguments of the case through its text. 
Each document enters the analysis in two textual forms, its full text and its summary, and we represent both as Sentence-BERT embeddings \citep{Reimers2019}: 384-dimensional numerical vectors in which documents with similar factual narratives or legal arguments lie close together. 
The two forms complement each other, since the full text retains every detail of a filing while the summary states its substance briefly and in a uniform format. 
The embedding of a document aggregates those of all documents filed in the case up to that point, so that it represents the record as a whole.

Because the vectors can be compared, they measure how similar any two documents, or any two cases, are. 
This makes it possible to do systematically what a litigator does from experience: to find the earlier cases that most resemble a filing and to see how they ended. 
For each filing, we retrieve its $k$ nearest neighbors, the $k$ previously resolved cases whose records are most similar to it, and record how often plaintiffs prevailed in them and how close they are. 
We use several values of $k$ ($10$, $50$, and $1{,}000$), so that the features describe both the closest comparators and the broader neighborhood of similar cases. 
A filing with many close comparators that plaintiffs mostly won thus carries a different signal from one with no close comparator at all. 
In addition, we include the embeddings themselves, compressed to eight dimensions with UMAP, so that the model also draws on a direct representation of the content of the record. 
All text-based features are computed separately from the full-text and the summary embeddings, which gives two parallel sets. 
Together, they capture the qualitative, case-by-case comparison on which legal assessments rest, a form of reasoning that until recently could not be measured systematically and at scale.

The third class captures the institutional actors. 
We identify the presiding judge and the law firms on each side, and record their experience and win rates in similar earlier cases. 
These features let the model draw on persistent patterns associated with particular judges and counsel, which prior work has documented \citep{Tippett2021,Mahari2024,Mojon2025}, without estimating a separate fixed effect for each. 
In total, each document is described by 98 features.

We code the outcome of each case from its final disposition as recorded in the filings. 
A plaintiff win is a judgment for the plaintiff. 
A plaintiff loss is a judgment for the defendant or a dismissal that ends the case, on the merits or on procedural grounds. 
A settlement is a disposition that the record identifies as a settlement.

\subsection{Predictive setup}

We treat every document as a point at which a forecast can be made. 
For each document, the model receives the features describing the case as it stood at that filing and returns a probability for each of the three outcomes. 
The predicted outcome is the one with the highest probability. 
A case with ten documents therefore yields ten successive forecasts, which allows us to follow how the assessment of a case changes as the record develops.

The prediction uses only information that a litigant could have had at the time, which three restrictions ensure. 
First, the information set of a case ends before its final disposition: we exclude the filing that records the disposition and every document filed on or after its date. 
Second, the comparison with similar cases and the experience and win rates of judges and law firms are computed exclusively from cases resolved before the focal case began. 
Third, we split the sample chronologically. 
The 80\% of cases that began earliest form the training set, and the 20\% that began latest form the test set, corresponding to approximately 20{,}000 cases and 140{,}000 documents. 
The split is made at the case level, so that no case contributes documents to both sets. 
The model is thus evaluated on later cases than those it learned from, as it would be used in practice.

Some features describe the procedural posture of the case, including rulings on motions to dismiss and for summary judgment. 
These rulings are part of the information a litigator would use to assess a case, and including them is intended: a partial grant of summary judgment, for example, changes the likely outcome of the case. 
What the model forecasts is how a case will end given its posture at the time of each filing.

We estimate the outcome probabilities with gradient-boosted decision trees \citep{Friedman2001}, a standard method for tabular data that captures non-linear relationships and interactions between features without requiring them to be specified in advance. 
We estimate two versions of the model: a baseline model that uses only the basic features, and a full model that uses all three feature classes. 
Unless stated otherwise, results refer to the full model, and Section~\ref{sec:richer} compares the two. 

Hyperparameters (tree depth, learning rate, and feature subsampling) are selected by three-fold cross-validation on the training set. 
Folds are grouped by case, so that documents from the same case never fall into different folds, and the selection criterion is the macro-averaged F1 score, which weights the three outcomes equally so that the most common one is not favored. 
The selected model is then retrained on the full training set. 
A random forest performs comparably and a ridge-regularized multinomial regression somewhat worse, so the results do not depend on a single model architecture. 
The full specification is given in \ref{app:model}, and further robustness analyses in \ref{app:robustness}.

\section{Results}
\label{sec:results}

We first report how predictable case outcomes are overall. 
The remaining subsections turn to the central question of the paper: which cases are hard to predict, and at what stage of the proceedings. 
For each filing, we measure outcome uncertainty as the entropy of the predicted distribution over the three outcomes, which we refer to as case complexity. 
We then examine how complexity varies across disputes, how much of it richer information removes, how it evolves as litigation proceeds, and how it relates to settlement.

\begin{figure}[htb]
    \centering
    \includegraphics[width=0.95\linewidth]{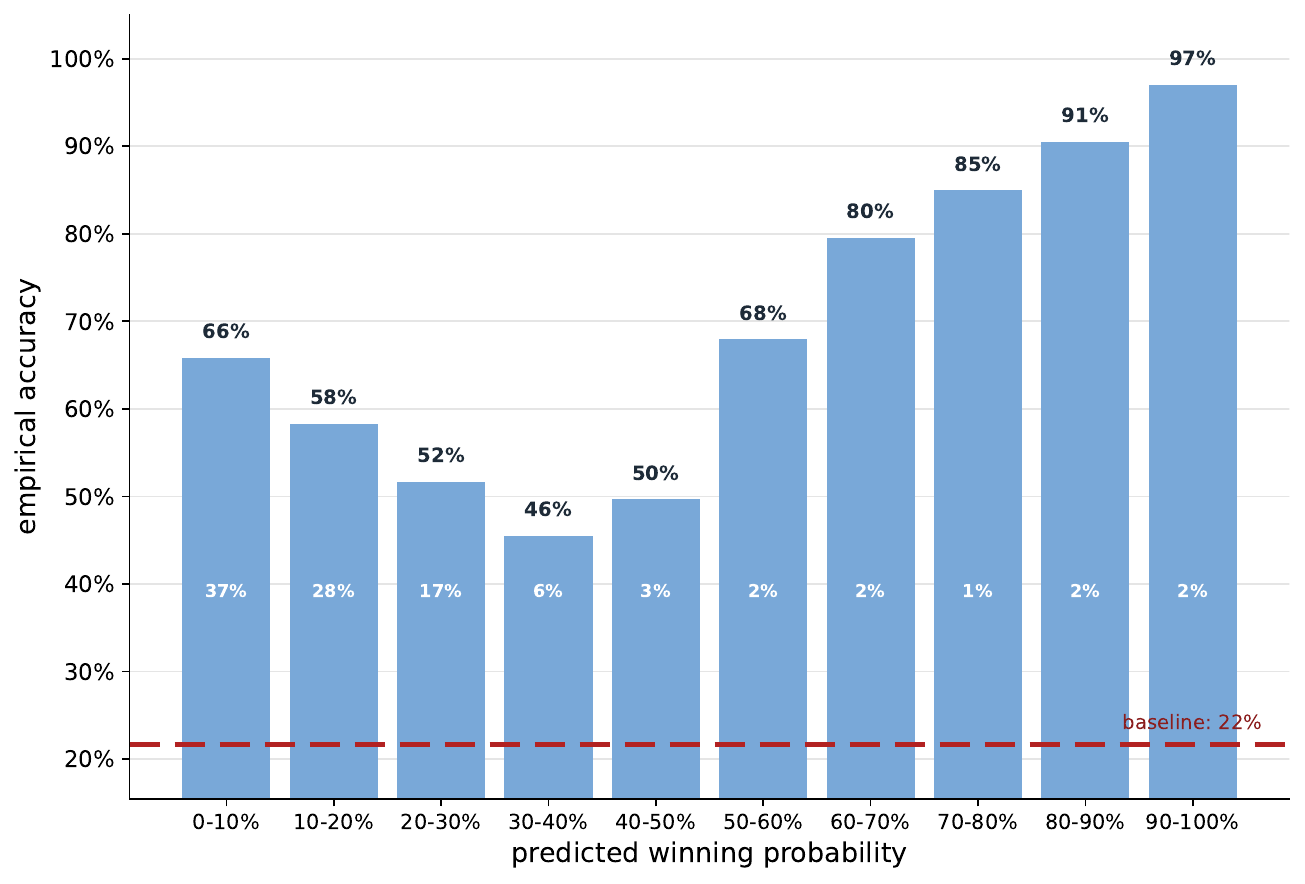}
	\caption{
	Prediction accuracy conditional on model confidence.
	Test documents are grouped into bins by the predicted probability of a plaintiff win.
	Bars show the share of documents whose outcome (plaintiff win, plaintiff loss, or settlement) is predicted correctly within each bin, and the dashed line indicates the share of plaintiff wins.
	Numbers inside the bars give the fraction of test documents in each bin.
	}
    \label{fig:accuracy_by_predicted_probability}
\end{figure}

\begin{figure*}[htb]
    \centering
    \includegraphics[width=\linewidth]{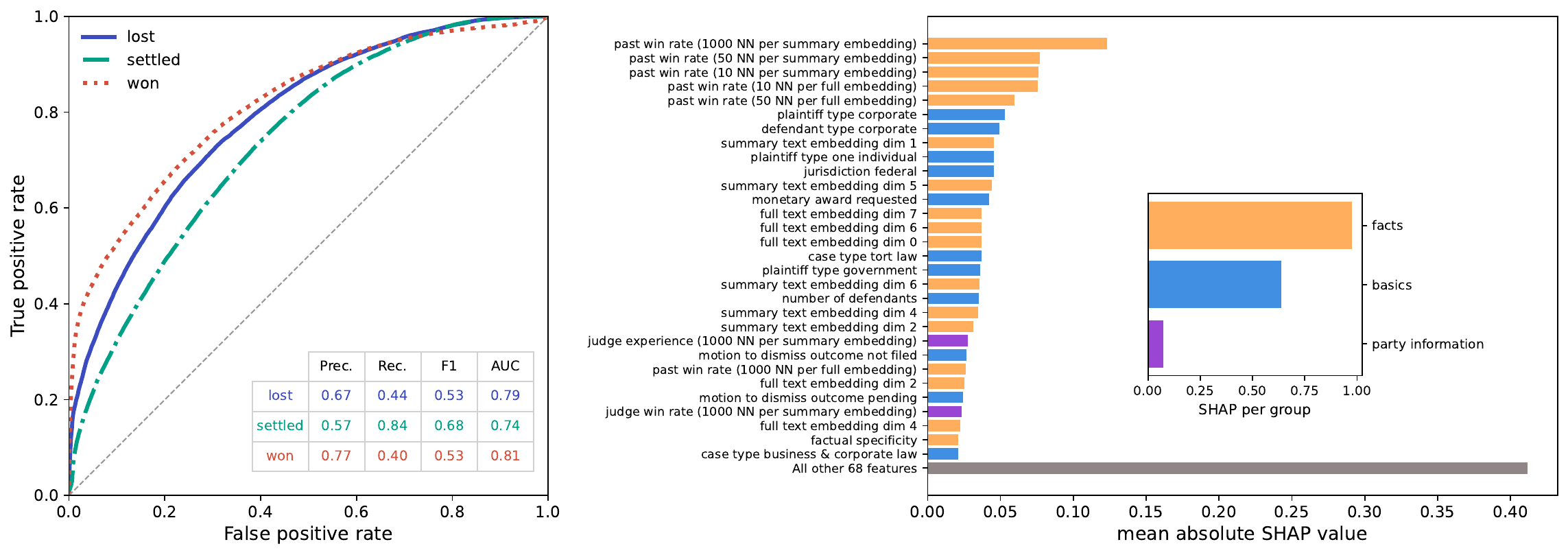}
	\caption{
	Baseline predictive performance and feature attribution. 
	(Left panel)
	Predictive performance for the three outcome classes (plaintiff win, plaintiff loss, and settlement), shown using receiver operating characteristic curves and summary classification metrics: precision, recall, F1-score and area under the curve (AUC). 
	(Right panel)
	Feature importance based on mean absolute SHAP values, with features ordered from most to least important. 
	The gradual decline in bar height indicates that predictive signal is distributed across many features rather than concentrated in a small number of dominant predictors. 
	Features are colored by feature group: basic structured variables, fact-based text features, and party-level information. 
	The inset reports the total SHAP contribution of each feature group.
	}
    \label{fig:predictability_per_class}
\end{figure*}

\subsection{Baseline predictive performance}

We first ask how predictable case outcomes are overall. 
Across the roughly 140{,}000 test documents, the full model predicts the correct one of the three outcomes for 61\% of documents. 
Always predicting the most common outcome in the test set, settlement, would be correct for 45\% of documents.

Average accuracy understates how predictable some cases are, because the model also reports how confident it is. 
Figure~\ref{fig:accuracy_by_predicted_probability} groups the test documents by the predicted probability of a plaintiff win. 
When that probability exceeds 90\%, the forecast is correct for 97\% of documents. 
Forecasts of a plaintiff loss are similarly reliable: when the predicted probability of a loss exceeds 90\%, 95\% of these forecasts are correct. 
Such confident forecasts of a win or a loss are made for about one document in twenty. 
In the middle of the probability range, accuracy falls to around one half. 
A small set of cases can thus be forecast with near certainty from the record, while for a large middle group the outcome remains uncertain.

The left panel of Figure~\ref{fig:predictability_per_class} reports performance for each outcome separately. 
The area under the receiver operating characteristic curve (AUC) is the probability that a randomly chosen document from a case with a given outcome receives a higher predicted probability of that outcome than a randomly chosen document from a case without it. 
A value of 0.5 corresponds to chance and a value of 1 to perfect separation. 
AUC values range from 0.74 for settlement to 0.81 for plaintiff wins. 
The model is cautious in predicting a plaintiff win: when it does, it is right 77\% of the time, but it identifies only 40\% of the documents from cases that plaintiffs won. 
Settlement is predicted more liberally, capturing 84\% of the documents from settled cases at a precision of 57\%. 
All three outcomes can thus be predicted to a substantial degree from information available before the disposition.

The right panel shows which features drive the predictions, measured by mean absolute SHAP values, which attribute each prediction additively to the individual features \citep{Chen2023}. 
The five most important features all record how similar earlier cases were resolved, namely the plaintiff win rate among the previously resolved cases whose documents most resemble the current one.\footnote{For instance, the feature ``past win rate (50 NN per summary embedding)'' refers to the average plaintiff win rate among the 50 previously resolved cases whose document summaries are most similar to the current document.} 
The strongest signal is thus a form of reasoning by analogy, close to how a litigator assesses a new matter against comparable cases. 
Party types, jurisdiction, and the amount of damages claimed follow. 
Beyond the top features, importance declines gradually, so the predictions draw on many features and not on a few. 
By feature class (inset), the facts of the case contribute most, followed by the basic features, while information on judges and law firms contributes comparatively little. 
This does not mean that judges and counsel do not matter, since their influence may already be reflected in the facts as pleaded and in the outcomes of similar cases. 
It means that, once the record and comparable cases are known, their identities add little predictive information.

\begin{figure}[htb]
    \centering
    \includegraphics[width=0.95\linewidth]{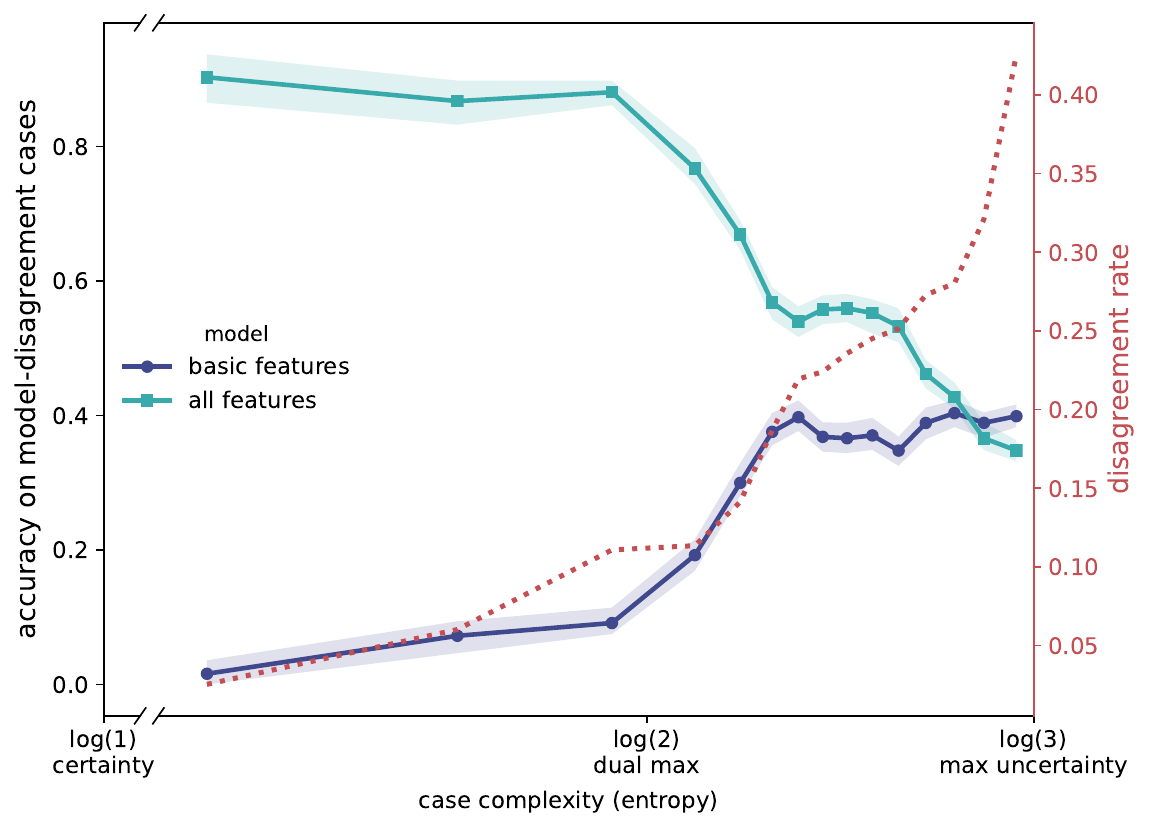}
	\caption{
	Model disagreement and accuracy by case complexity. 
	Test documents are ordered by the entropy of the full model's predicted outcome distribution and divided into fifteen equally sized groups of increasing complexity. 
	(Right axis)
	Share of documents for which the baseline model (basic features only) and the full model (all features) predict different outcomes. 
	(Left axis)
	Accuracy of each model on the documents for which the two models disagree. 
	The full model is right far more often in low-complexity cases, and its advantage disappears as complexity increases.
	}
    \label{fig:excess_predictability}
\end{figure}

\subsection{Richer features help most in simple cases}
\label{sec:richer}

The facts of the case contribute most to the predictions, yet in aggregate they improve them only modestly. 
A baseline model that uses only the basic features reaches AUC values of 0.75 (loss), 0.70 (settlement), and 0.79 (win), and the full model reaches 0.79, 0.74, and 0.81. 
Overall accuracy rises from 58\% to 61\%. 
Aggregate measures, however, average over very different cases, and whether richer information helps may depend on how hard a case is to begin with.

To make this precise, we measure case complexity by predictive uncertainty. 
For each document, the full model produces a probability distribution over the three outcomes, and we define complexity as the entropy of this distribution, $H = -\sum_i p_i \log p_i$, where $i$ indexes the three outcomes.\footnote{A distribution concentrated on a single outcome, such as $(1,0,0)$, yields zero entropy ($\log(1)=0$) and corresponds to a highly predictable case. A uniform distribution, $(1/3,1/3,1/3)$, yields the maximum entropy in this three-class setting ($\log(3)\approx 1.10$) and corresponds to maximal uncertainty. Intermediate cases fall between these extremes; for instance, $(1/2,1/2,0)$ corresponds to maximal uncertainty across two outcomes ($\log(2)\approx 0.69$), while $(0.7,0.2,0.1)$ yields a moderate entropy level ($\approx 0.80$).} 
Low entropy means that the model concentrates its probability on one outcome, and high entropy that it spreads probability across several. 
In the terms of Section~\ref{sec:literature}, low-entropy cases are easy cases and high-entropy cases are hard ones, with the qualification that both are judged by what the record reveals to the model, not by what the court or the parties know. 
We use the term case complexity in this sense throughout.

Figure~\ref{fig:excess_predictability} orders the test documents by entropy and divides them into fifteen equally sized groups. 
Within each group, it shows how often the baseline and full models predict different outcomes (right axis) and, for the documents on which they disagree, how often each model is right (left axis). 
Disagreement rises steadily with complexity, from 3\% of documents in the lowest group to 45\% in the highest. 
The more uncertain a case, the more the forecast depends on the information used.

When the two models disagree on a low-complexity case, the full model is almost always right. 
In the lowest group, it is correct for 90\% of the disagreements and the baseline model for 2\%. 
As complexity rises, this advantage shrinks, and in the most complex groups the two models are right about equally often, with the baseline model slightly ahead in the highest group. 
Accuracy over all documents in a group shows the same pattern: the full model gains up to nine percentage points in groups of low and intermediate complexity and none in the most complex. 
Richer information about the facts, comparable cases, judges, and counsel thus sharpens forecasts for cases that are already relatively clear, but does little for the hardest cases. 
A subset of disputes therefore remains hard to predict even when considerably more observable information is supplied. 
Whether this reflects legal indeterminacy or information that is not in the record, such as private information held by the parties, cannot be determined from these data.

\begin{table}[!htbp]\centering
\scriptsize
\setlength{\tabcolsep}{3pt}
\sisetup{table-number-alignment=center, input-symbols=()}
\caption{
	Observable case characteristics explain variation in case complexity. 
	The table reports coefficients from a linear regression of entropy on case-level and document-level features. 
	Positive coefficients indicate higher predictive uncertainty (greater complexity), while negative coefficients indicate more predictable cases. 
	}
\begin{tabular*}{\columnwidth}{@{\extracolsep{\fill}}lS[table-format=-1.3]@{}lS[table-format=1.3]@{}}
\toprule
 & \multicolumn{2}{c}{Coef.} & {SE} \\
\midrule
Jurisdiction: state 						& -0.213 & *** & 0.034 \\
Case type: business \& corporate law 		& 0.271 & ** & 0.107 \\
Case type: contract law 					& -0.071 &  & 0.107 \\
Case type: employment law 				& 0.028 &  & 0.103 \\
Case type: intellectual property law			& 0.325 & *** & 0.102 \\
Case type: property \& real estate law 		& 0.372 & *** & 0.107 \\
Case type: tort law 						& -0.693 & *** & 0.101 \\
Plaintiff type: government 					& -0.224 & *** & 0.081 \\
Plaintiff type: multiple individuals 			& -0.258 & *** & 0.055 \\
Plaintiff type: one individual 				& -0.281 & *** & 0.025 \\
Defendant type: government 				& -0.137 & *** & 0.046 \\
Defendant type: multiple individuals 			& -0.151 & *** & 0.029 \\
Defendant type: one individual 				& -0.214 & *** & 0.040 \\
Number of plaintiffs						& 0.003 & *** & 0.001 \\
Number of defendants 					& -0.051 & *** & 0.014 \\
Past win rate (10 NN per summary embedding) & -0.346 & *** & 0.015 \\
Duration 								& 0.081 & *** & 0.010 \\
\midrule
Observations & \multicolumn{3}{r}{138,771} \\
$R^2$ & \multicolumn{3}{r}{0.35} \\
\bottomrule
\multicolumn{4}{p{0.96\columnwidth}}{
\scriptsize \textit{Notes:} 
The dependent variable is standardized entropy from the all-feature model. 
Non-fixed-effect covariates are standardized. 
Standard errors are clustered by case ID. 
Reference categories are 
Jurisdiction: federal; 
Case type: bankruptcy law; 
Plaintiff type: corporate;
Defendant type: corporate. 
$^{***}p<0.01$, $^{**}p<0.05$, $^{*}p<0.10$.
} \\
\end{tabular*}
\label{tbl:complexity_regression} 
\end{table}

\subsection{Complexity varies with area of law, parties, and comparable cases}

We next ask which observable characteristics of a case are associated with its complexity. 
Table~\ref{tbl:complexity_regression} reports a linear regression of standardized entropy on jurisdiction, the area of law, the types and numbers of parties, the time elapsed since the case began, and the plaintiff win rate among comparable earlier cases, with standard errors clustered by case. 
The regression is descriptive: its coefficients show where uncertainty is concentrated, not what causes it. 
These characteristics explain 35\% of the variation in complexity across 138{,}771 test documents, so the uncertainty that remains after prediction is far from random.

Complexity differs sharply across areas of law. 
Tort cases are the most predictable, with entropy 0.69 standard deviations below that of bankruptcy cases, the reference category. 
Business and corporate, intellectual property, and property and real estate disputes are the least predictable, while contract and employment cases do not differ significantly from bankruptcy. 
Cases in state courts are more predictable than cases in federal courts, consistent with the large share of routine claims on state dockets \citep{HannafordAgor2015}.

Party configuration matters as well. 
Cases with a corporate plaintiff or a corporate defendant are the hardest to predict: relative to corporate parties, individuals, groups of individuals, and government entities on either side are all associated with lower entropy. 
Corporate litigants may bring more complex claims, have more at stake, or litigate more strategically, and as repeat players they can choose which disputes to pursue to judgment \citep{Galanter1974}. 
The number of parties matters less. 
More defendants are associated with slightly lower complexity, and the coefficient on the number of plaintiffs, although statistically significant, is negligible.

The plaintiff win rate among the ten most similar previously resolved cases is the strongest continuous correlate of complexity. 
A one-standard-deviation increase in this rate is associated with entropy 0.35 standard deviations lower. 
Cases whose closest comparators were mostly won by plaintiffs are therefore easier to forecast than cases whose comparators were mostly lost or settled. 
One reading is that a low win rate among comparable cases tells the model that the plaintiff is unlikely to prevail, but not whether the case will be lost or settled, so the predicted probability is split between those two outcomes.

Finally, complexity is higher for documents filed later in a case.\footnote{Duration is measured as the elapsed time between the start of the case and the filing date of the current document.} 
Because the regression pools documents across cases, this coefficient can reflect either a change within cases or the fact that only long cases contribute late documents. 
The next section shows that it reflects the latter.

\begin{figure}[htb]
    \centering
    \includegraphics[width=0.95\linewidth]{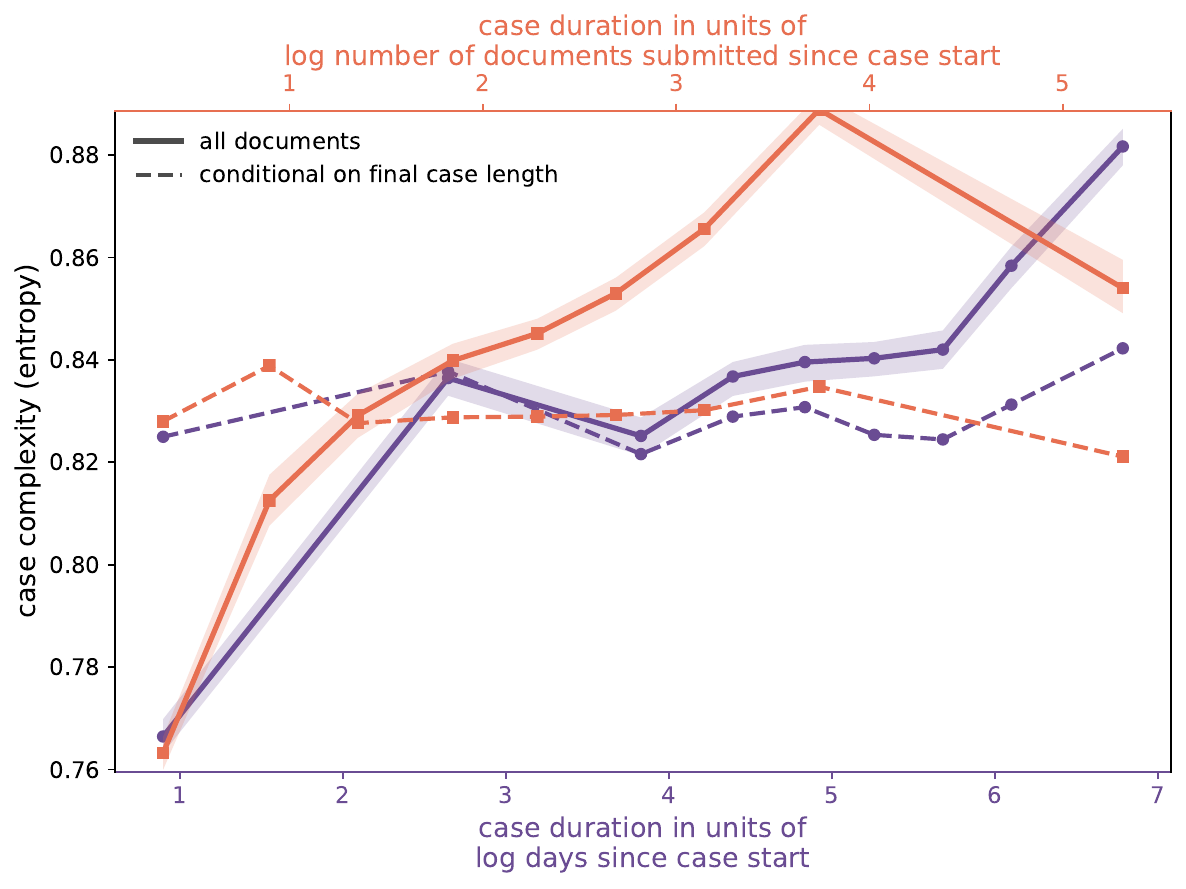}
	\caption{
	Harder cases last longer. 
	Average entropy of the predicted outcome distribution for test documents, by stage of the case. 
	The bottom axis measures the time elapsed since the case began (log days), and the top axis the number of documents filed so far (log scale). 
	Solid lines pool all documents, with shaded bands showing 90\% bootstrap intervals. 
	Dashed lines show the same relation after conditioning on the final length of each case, measured by its number of documents. 
	}
    \label{fig:case_progression}
\end{figure}

\subsection{Harder cases last longer}
\label{sec:progression}

The positive association between duration and complexity raises the question of how predictability evolves over the life of a case. 
Figure~\ref{fig:case_progression} plots average complexity against two measures of progress: the time elapsed since the case began (bottom axis, log days) and the number of documents filed so far (top axis, log scale).

Pooled across cases (solid lines), complexity rises as litigation proceeds. 
On the time axis, average entropy increases from about 0.77 among the earliest documents to about 0.88 among the latest. 
Late documents, however, come only from cases that last long, so the pooled curve mixes two effects: cases may become harder to predict as they proceed, or long cases may be harder to predict from the outset. 
The dashed lines separate the two by showing the same relation after conditioning on the final length of each case, measured by its number of documents. 
Once case length is held fixed, the increase disappears, and average entropy stays between about 0.82 and 0.84 on both axes.

The rise in the pooled curve therefore reflects which cases are still pending, not a change within cases. 
Cases that will last long are already harder to predict at their first filing, and within cases of a given length the forecast does not change systematically as the record grows. 
Two conclusions follow. 
First, harder cases last longer. 
Cases that are easy to predict tend to end early, through settlement, default, or dismissal, so the cases still pending at later stages are increasingly those whose outcome was uncertain from the start. 
This is a dynamic counterpart of the selection of disputes for trial described by \citet{Priest1984}. 
Second, additional filings do not resolve the uncertainty. 
As pleadings are answered, motions are decided, and the record develops, one might expect forecasts to sharpen, yet on average they do not, even though the model sees every new document. 
This parallels the finding in Section~\ref{sec:richer} that richer information does little for hard cases.

\subsection{Settlement rates peak at intermediate complexity}
\label{sec:settlement}

\begin{figure}[htb]
    \centering
    \includegraphics[width=0.95\linewidth]{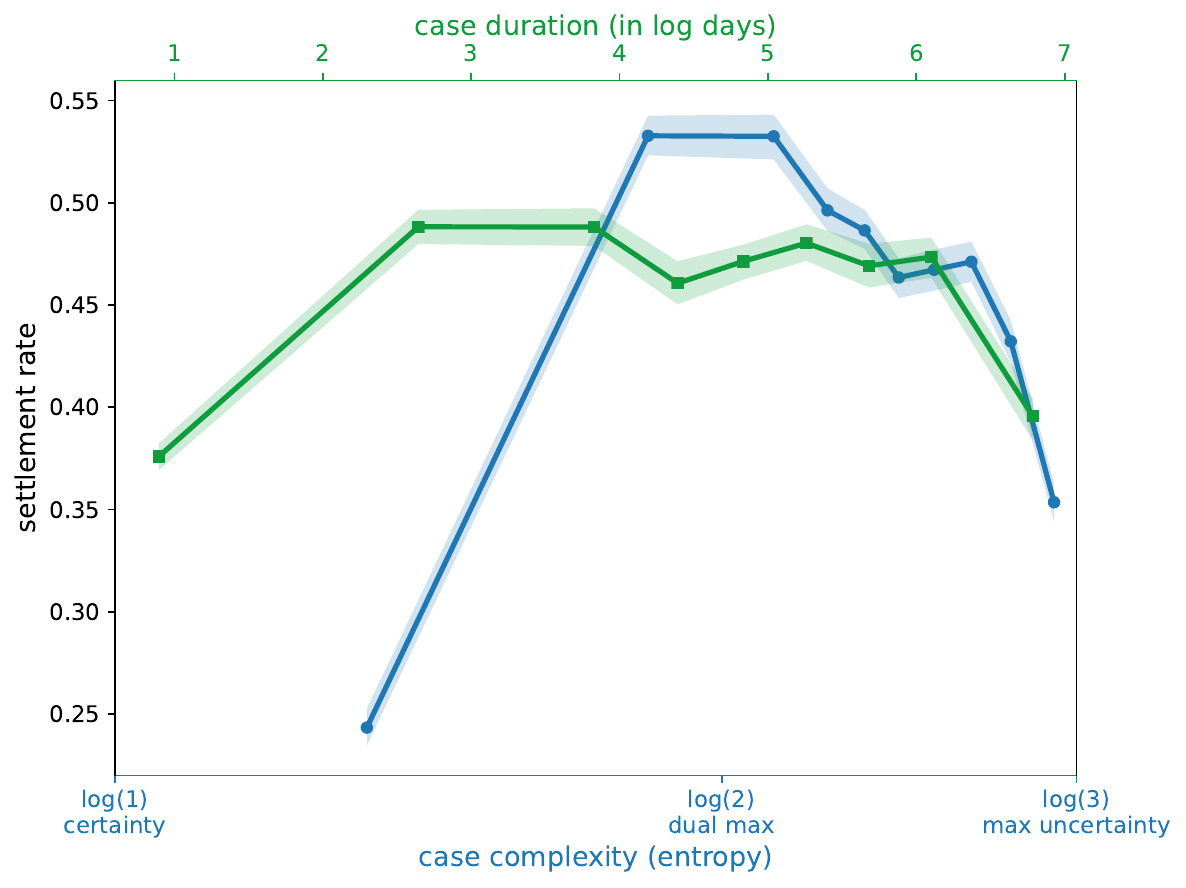}
	\caption{
	Settlement rates exhibit an inverted U-shape with respect to case complexity. 
	Share of test documents from cases that eventually settled, by decile of entropy (bottom axis) and by decile of the time elapsed since the case began (top axis, log days). 
	Settlement is least common among highly predictable cases, most common at intermediate complexity, and less common again among the most complex cases. 
	}
    \label{fig:settlement_rate}
\end{figure}

Economic models of litigation make settlement depend on the parties' uncertainty about the outcome (Section~\ref{sec:literature}). 
We therefore ask how settlement relates to case complexity. 
Figure~\ref{fig:settlement_rate} plots the share of test documents from cases that eventually settled against complexity (bottom axis) and against the time elapsed since the case began (top axis, log days).

The settlement rate follows an inverted U. 
Among the least complex documents, about a quarter come from cases that settled. 
The rate rises to more than half at low to intermediate complexity and falls to about a third among the most complex documents.\footnote{A similar pattern is observed when using duration as the horizontal axis, reflecting the association between case length and complexity documented above.} 
Clear cases and highly uncertain cases are thus both resolved by adjudication more often than the cases in between.

Clear cases may end in adjudication because there is little to bargain over: a defendant facing a near-certain loss may default, and a weak claim may be dismissed. 
Highly uncertain cases may fail to settle because the parties' expectations diverge, as in the model of \citet{Priest1984}. 
Because entropy is computed from the predicted probabilities, including that of settlement, part of the pattern follows from the measure itself. 
A forecast close to certainty is in most cases a forecast of a win or a loss, whereas a maximally uncertain forecast assigns settlement a probability of about one third. 
The figure therefore shows where settlements are concentrated along the scale of uncertainty, and the explanations above remain hypotheses.

\section{Discussion}

This article provides a large-scale measurement of how predictable the outcomes of civil lawsuits are, using only what was on the record at each point in a case. 
The forecasts rest largely on comparing each filing with the most similar earlier cases and how they were resolved, a comparison that practitioners make from experience and that language models now allow to be made systematically from the full text of the record. 
We report five findings. 
Outcomes are predictable to a substantial degree, and for a small share of cases nearly certain. 
The uncertainty that remains is concentrated in identifiable areas of law, jurisdictions, party configurations, and kinds of cases. 
Richer information sharpens forecasts for clear cases but does little for hard ones. 
Harder cases last longer, and additional filings do not make their outcome clearer. 
And settlement is most common at intermediate levels of uncertainty. 
We discuss these findings in turn and then consider their implications for legal theory and practice.

\subsection{Prediction under evolving legal information}

Across three outcomes, forecasts made from the record are correct for about three in five documents, against 45\% for always predicting the most common outcome, and the most confident forecasts of a win or a loss are correct for 95 to 97\% of documents. 
Accuracy figures are not directly comparable across studies, because most prior work predicts a binary result, often for a single apex court or a single area of law, and often from information that becomes available only with the decision \citep{Ruger2004,Katz2017,Alexander2019,Mcconnell2021,Medvedeva2023}. 
Our estimates are conservative in this respect: they use only filings made before the outcome, include settlement as a third outcome, and are evaluated on later cases than those the model learned from. 
They extend the evidence on predictability from apex courts to ordinary civil litigation in state and federal courts.

The confident forecasts also matter in practice. 
Lawyers' assessments of their own cases are often poorly calibrated, and parties who reject settlement offers frequently do worse at trial \citep{GoodmanDelahunty2010,Kiser2008}. 
By contrast, when the model assigns a probability above 90\% to a win or a loss, that forecast is borne out in nearly all cases. 
Such forecasts are available for only about one document in twenty, but they identify a set of disputes whose outcome can be anticipated with near certainty from the record.

The sources of predictability are consistent with how lawyers reason. 
The strongest signal is the outcome of comparable earlier cases, a statistical counterpart of reasoning by analogy, followed by the area of law and the types of parties \citep{Ashley2009,Cao2024,Valvoda2024}. 
Prior work shows that the identity of judges and counsel can predict outcomes \citep{Chen2017,Tippett2021,Mahari2024,Mojon2025}. 
In our data, information on judges and law firms adds little once the record and comparable cases are known. 
This does not show that judges and counsel are unimportant, since their influence may already be reflected in the pleadings and in the outcomes of similar cases. 
It suggests, however, that for the purpose of forecasting much of their effect is captured by the case itself.

\subsection{Case complexity and legal indeterminacy}

Predictability in civil litigation is highly heterogeneous. 
Some cases can be forecast with near certainty from the record, while others remain uncertain even when considerably more information is supplied. 
This maps onto the jurisprudential distinction between easy cases, in which the governing law applied to the facts settles the result, and hard cases, in which legal sources underdetermine it \citep{Dworkin1975,Schauer1985,Marmor1990}. 
Our entropy measure gives this distinction an empirical counterpart for ordinary civil litigation, in the spirit of using calibrated forecasts to measure how strongly the law guides outcomes \citep{Stiglitz2026}.

Two features of the results speak to the debate over how common easy cases are. 
First, near-certain forecasts are available for only a small share of documents. 
Judges tend to describe most cases as easy \citep{Posner2010,Howard2014}, whereas selection models predict that easy disputes settle or are never filed, leaving the hard ones in court \citep{Priest1984,Fischman2021}. 
Our measure cannot resolve this question, because the model's uncertainty combines legal indeterminacy with information that is missing from the record. 
The share of near-certain cases is therefore best read as a lower bound on the share of cases that are easy in the legal sense. 
Even so, it suggests that among filed civil cases, outcomes that are clear from the record are the exception and not the rule.

Second, complexity is concentrated in identifiable parts of the docket. 
Business and corporate, intellectual property, and property disputes, as well as cases with corporate parties, are the least predictable, while tort cases, cases with individual or government parties, and cases in state courts are the most predictable. 
This is compatible with the view that indeterminacy is moderate and unevenly distributed, not pervasive \citep{Solum1987,Kress1989}, and with recent evidence that the difficulty of cases varies systematically \citep{Sheppard2026}.
The ranking of tort cases differs from that of \citet{Waldfogel1995}, who infers from trial and win rates that outcome uncertainty is higher in tort cases than in contract and property cases. 
The two measures capture different quantities: his estimates concern the parties' uncertainty in the cases that reach trial, whereas ours measure the uncertainty of forecasts made from the record across all filed cases, including those that settle or end without trial.

Finally, richer information sharpens forecasts for clear cases but adds little for the hardest ones. 
This is what one would expect if some disputes are underdetermined by law. 
It is equally consistent, however, with uncertainty that stems from information outside the record, such as the parties' private information, the strength of the evidence, or the conduct of trial. 
Distinguishing these sources would require data that link forecasts to the legal questions at issue in each case, and we leave it to future work.

\subsection{The dynamics of complexity}

Complexity is higher among documents filed later in litigation, but this reflects which cases are still pending, not a change within cases. 
Holding the final length of a case fixed, complexity does not change systematically as filings accumulate. 
We consider the result against two views of how litigation unfolds.

The first is litigation as the revelation of information, in which each new filing narrows the range of plausible outcomes and motion practice in particular moves the parties toward agreement \citep{Spier1992,Boyd2013}. 
Our forecasts, which are made from the public record, do not on average become sharper as the record grows. 
This does not mean that the parties learn nothing, since they hold private information and may update on events that the filings do not capture. 
It does suggest that the public record of a hard case rarely resolves what made it hard.

The second is selection over the life of a case. 
Cases with clear outcomes tend to end early, through settlement, default, or dismissal, and the cases still pending at later stages are those whose outcome was uncertain from the outset. 
This is a dynamic analogue of the selection of disputes for trial \citep{Priest1984}, operating at every stage of a case and not only at the choice between settlement and trial. 
Our results are consistent with this view: the cases that remain are hard from the start, and additional filings mainly reveal which cases these are. 
Because litigation costs are incurred in stages, such cases can be pursued over many stages \citep{Bebchuk1996}, and uncertainty about how long they will last adds to the cost of doing so \citep{Lera2026}.

\subsection{Settlement and intermediate uncertainty}

Settlement is most common at intermediate levels of complexity and less common both where the outcome is clear and where it is highly uncertain. 
The second half of this pattern fits standard theory, but the first does not. 
In the divergent-expectations model, settlement fails when the parties' estimates of the outcome differ by more than the cost of litigating, which is most likely when the outcome is uncertain \citep{Priest1984}. 
Private information and self-serving assessments of the merits widen that gap further \citep{Bebchuk1984,Loewenstein1993}. 
The same models predict that cases with clear outcomes should settle most readily, since the parties have little to disagree about.

In our data, clear cases are instead often resolved by adjudication. 
A plausible reason is that many of them are never contested in substance. 
When a defendant faces a near-certain loss, the case may end in default or in a judgment that is not seriously opposed, and a plainly weak claim may be dismissed on motion. 
Much of state civil litigation consists of such routine matters, frequently against unrepresented defendants \citep{HannafordAgor2015,WilfTownsend2022}, and in federal courts adjudication without trial accounts for much of what is often attributed to settlement \citep{Hadfield2004}. 
Bargaining models presuppose two parties who litigate. 
Where one side does not, adjudication without trial, not settlement, is the natural end of a clear case.

At intermediate complexity, both parties face meaningful litigation risk, which widens the range of mutually acceptable terms and makes settlement attractive, consistent with models of pretrial bargaining under uncertainty \citep{Spier1992,Spier2007}. 
These results reinforce the view that settlement is not a residual category but a disposition whose frequency varies systematically with the characteristics of a case \citep{Galanter1993,Eisenberg2009}. 
Two qualifications apply. 
Part of the inverted U follows from the construction of the entropy measure, as noted in Section~\ref{sec:settlement}, and our outcome categories do not separate adjudication at trial from adjudication without trial. 
Three extensions would show which mechanisms drive the pattern. 
First, an uncertainty measure that excludes settlement, such as the uncertainty of a forecast between a plaintiff win and a plaintiff loss, would remove the mechanical component. 
Second, distinguishing defaults, dismissals, summary judgments, and trials would test whether clear cases end in adjudication because they are never contested. 
Third, relating the uncertainty of a case at each stage to whether it settles at the next would show when in the proceedings settlement responds to uncertainty. 
A bargaining model that combines the parties' private information with a public signal of case strength, litigation costs incurred in stages, and the option of one side not to contest would make these predictions precise. 
In such a model, settlement is rare when the public signal is clear, because the weaker side does not contest, and when it is weak, because expectations diverge; the model would also predict how the peak shifts with the stakes, litigation costs, and third-party funding.

\subsection{Implications}

For legal theory and empirical legal research, the main implication is that predictability can be measured case by case. 
Debates about legal indeterminacy and about how many cases are easy have relied on argument, appellate votes, or judges' own reports \citep{Schauer1985,Fischman2021}. 
Calibrated forecasts made from the record offer a complementary measure that covers ordinary civil cases, including those that settle \citep{Stiglitz2026}. 
They can also serve the literature on the selection of disputes. 
Litigation models are defined in terms of the parties' beliefs about the plaintiff's chance of winning \citep{Gelbach2018}, but empirical tests have had to infer those beliefs from trial rates and win rates \citep{Klerman2014}. 
Forecasts made at each stage of a case are an observable, if imperfect, proxy. 
They are not the parties' beliefs, but they show what could be inferred from the public record at the time.

For practice, the results suggest that forecasts should be read together with their uncertainty. 
The model is most reliable in the tails of its predicted distribution, so confident forecasts may be useful for case triage, settlement valuation, reserving by insurers, and the pricing of claims by litigation funders \citep{Katz2013,Lera2026}. 
In high-entropy cases, by contrast, the uncertainty is itself the relevant output. 
It identifies disputes in which the record does not determine the outcome and in which legal judgment, not statistical prediction, has to carry the analysis. 
The settlement results add that uncertainty is related to how a case is likely to end, not only to who is likely to prevail.

Calibrated forecasts may also make litigation more efficient. 
In the economic analysis of settlement, cases fail to settle mainly because the parties' expectations about the outcome diverge, and self-serving assessments widen the gap \citep{Priest1984,Loewenstein1993}. 
Forecasts that both sides can consult narrow that gap and could bring earlier resolution at lower cost. 
For the same reason, they matter for litigation funding: funders price claims under uncertainty, and more reliable assessments of case strength could lower the cost of financing meritorious claims that plaintiffs could not otherwise pursue \citep{Lera2026}. 
Our results also indicate the limits of these gains. 
Forecasts improve most for cases that are already relatively clear, so the disputes in which uncertainty is greatest, and in which parties most need help in assessing their position, are those in which prediction helps least.

These benefits depend on who uses the forecasts, which raises a distributional question for procedure and access to justice. 
Outcome-prediction tools shift information and costs among litigants \citep{Engstrom2021}, and repeat players are best placed to use them \citep{Galanter1974}. 
If forecasts are most accurate for routine cases, and routine litigation on state dockets often pits high-volume corporate plaintiffs against unrepresented individuals \citep{WilfTownsend2022}, prediction may strengthen the position of the party that already holds the advantage. 
Making such tools available to both sides, as has been proposed for legal research \citep{Tippett2021}, is one response.

These points caution against treating prediction as a substitute for legal reasoning. 
The value of such models lies less in classifying outcomes than in estimating risk, locating uncertainty, and distinguishing cases in which a forecast is informative from those in which the limits of predictability are themselves substantively meaningful.

\section{Conclusion}

Lawyers, parties, insurers, and litigation funders act every day on forecasts of how civil cases will end, yet the predictability of litigation has rarely been measured. 
Using more than 100{,}000 civil cases and the filings in them, we provide a first large-scale measurement. 
Using only what was on the record at the time, forecasts clearly beat a naive benchmark, and for a small share of cases they are nearly certain. 
The uncertainty that remains is not random. 
It is concentrated in particular areas of law, jurisdictions, and party configurations, it is reduced little by richer information in the hardest cases, and it is related to how often disputes settle. 
Cases that are hard to predict at their first filing stay hard to predict and last longer. 
Treating the entropy of calibrated forecasts as a measure of case complexity turns the distinction between easy and hard cases into a quantity that can be estimated for every case.

Two limitations qualify these results. 
First, our measure captures uncertainty as it appears from the public record. 
It combines legal indeterminacy with information the record does not contain, such as the parties' private assessments, negotiations, and evidence that is never filed, and it identifies where uncertainty is concentrated rather than what causes it. 
Because a richer information set could only sharpen the forecasts, the predictability we report is best read as a lower bound.

Second, outcomes are coded from the dispositions recorded in the filings. 
Settlements that enter the docket only as a dismissal are therefore coded as plaintiff losses, and the sample is limited to filed cases that reached a recorded final disposition, so disputes resolved before filing and cases still pending fall outside it.

Several extensions follow. 
The first is to separate legal indeterminacy from information missing from the record, for example by linking forecasts to the legal questions at issue, the procedural events that resolve them, and the way each case was disposed of. 
The second is to explain why settlement peaks at intermediate uncertainty, by combining an uncertainty measure that excludes settlement and data on how non-settled cases end with an economic model of settlement in which the parties' incentives, costs, and information can be varied. 
The third is to apply the approach across jurisdictions and over time, which would allow comparative study of how strongly the law guides outcomes in different courts and areas of law. 
Such work would move the study of legal predictability from individual courts and areas of law toward systematic measurement across the civil justice system.

\bibliographystyle{elsarticle-harv}
\bibliography{bibliography}
\balance
\clearpage
\appendix
\onecolumn

\section{Data Processing and Outcome Coding}
\label{app:processing}

\noindent\textbf{Source and sample.}
The data consist of the dockets and documents of 102{,}721 civil cases filed between 1996 and 2022, randomly sampled from state and federal courts in 12 U.S.\ states (Arizona, California, Delaware, Florida, Illinois, Massachusetts, New Jersey, New York, Ohio, Pennsylvania, Texas, and Virginia). 
The records originate from the federal judiciary's PACER system and from the electronic records of state courts and were obtained through commercial providers of court records. 
Each case consists of its docket, which lists the filings in the case, and the documents attached to the docket entries, available as PDF files. 
Every document is linked to its case and carries its filing date and docket number, so that the documents of a case can be ordered as they were filed; documents without a valid filing date are dropped, and where two documents share a docket number, only the first is kept.

\noindent\textbf{Text extraction.}
All documents are converted from PDF to text by optical character recognition.
The text is cleaned of extraneous whitespace and non-textual characters. 
For the language model, documents that exceed its input limit are shortened by keeping their beginning and their end, where the caption, the relief sought, and any ruling are usually stated. 
Administrative and service filings and exhibits are not used, since they rarely contain information about the merits of a case.

\noindent\textbf{Information extraction.}
Each document is passed to a large language model with a structured prompt that asks a fixed set of questions and requires the answers in a fixed format, with an explicit option to indicate that the document does not contain the answer. 
The questions cover the area of law, the type of document and the filing party, the number and type of parties, the relief and damages sought, references to evidence, the status of motions to dismiss and for summary judgment, the posture of the case, and, for final dispositions, the prevailing party. 
The model also writes a summary of the document of at most 500 characters. 
\ref{app:features} lists the resulting variables. 
Case-level attributes that a document does not state are carried forward from earlier documents of the same case, and categorical variables are one-hot encoded.

\noindent\textbf{Text representations.}
We compute 384-dimensional Sentence-BERT embeddings \citep{Reimers2019} of each summary and of the full text of each document, processing long documents in chunks whose embeddings are aggregated. 
The embeddings are cumulative: the representation of the $\ell$-th document of a case aggregates the embeddings of all documents up to and including it, so that it reflects the whole record at that point. 
For use as model inputs, each embedding is compressed to eight dimensions with UMAP.
The summary and full-text representations are processed in parallel and yield two sets of text features. 
The summaries are short and follow a uniform format, so they capture the substance of a filing with little noise from formatting, boilerplate, or errors in text recognition; the full text retains details that a 500-character summary leaves out. 
The model receives both sets: eight compressed embedding dimensions from each representation, and the comparable-case features described below, computed separately in each embedding space.

\noindent\textbf{Comparable cases.}
For each filing, we retrieve the $k$ previously resolved cases whose records are most similar to it ($k = 10$, $50$, and $1{,}000$), measured by cosine distance in the full embedding space, with each earlier case represented by the embedding of its final document. 
We record the plaintiff win rate among these cases and their minimum and mean distance to the current filing. 
With three values of $k$ and two embedding spaces, this yields eighteen comparable-case features. 
The experience and win rates of law firms are computed in the same way from the 1{,}000 most similar earlier cases in each embedding space, and those of judges from the 1{,}000 most similar earlier cases in the summary space. 
Only cases resolved before the focal case began enter these comparisons.

\noindent\textbf{Outcome coding.}
For every document, the language model records the posture of the case, whether further proceedings are expected, and, if the document records a final disposition, which party prevailed. 
A document counts as disposing of the case if it records a judgment, a dismissal with prejudice, or a settlement, and no further proceedings are expected. 
The outcome of the case is the disposition on which at least 60\% of these documents agree. 
Cases without such agreement, and cases without a disposing document, are excluded. 
Judgments for the plaintiff are coded as plaintiff wins, judgments for the defendant and dismissals as plaintiff losses, and recorded settlements as settlements. 
The case ends on the date of the first disposing document, and that document and all documents filed on or after its date are excluded from the information set.

The resulting settlement share of 34\% is lower than the roughly two-thirds that \citet{Eisenberg2009} estimate for two federal districts. 
The difference reflects how settlements appear in court records. 
Many settlements are concluded privately and reach the docket only as a voluntary dismissal, which the record does not distinguish from other dismissals \citep{Hadfield2004,Eisenberg2009}. 
Our settlement category therefore captures settlements that appear in the record as such, and privately settled cases that end in a dismissal are coded as plaintiff losses.

\section{List of Features}
\label{app:features}

This appendix provides additional detail on the feature set used in the predictive models.  
The underlying dataset comprises 102{,}721 civil cases and 835{,}190 court-filed PDF documents spanning 12 U.S.\ states: Arizona, California, Delaware, Florida, Illinois, Massachusetts, New Jersey, New York, Ohio, Pennsylvania, Texas, and Virginia.  

The predictive model contains 98 encoded features after categorical variables are one-hot encoded.  
Table~\ref{tab:feature_appendix} reports the features in grouped form for readability; as a result, some rows correspond to multiple encoded model variables, such as embedding dimensions, nearest-neighbor statistics, or one-hot indicators derived from categorical variables.  

The unit of prediction is the document.  
Most features are therefore document-level variables measured at the time of the current filing.  
A smaller set of variables captures broader case characteristics, including jurisdiction, case type, plaintiff type, defendant type, and the number of plaintiffs and defendants.  
All time-varying features are constructed using only information available up to the current document.  

\begin{longtable}{p{0.18\textwidth}p{0.31\textwidth}p{0.43\textwidth}}
\caption{Feature groups used in the predictive model.}
\label{tab:feature_appendix} \\

\toprule
Feature group & Feature & Description \\
\midrule
\endfirsthead

\toprule
Feature group & Feature & Description \\
\midrule
\endhead

\bottomrule
\endfoot

Basics 
& Jurisdiction 
& Court system in which the case is filed, classified as federal or state. \\

Basics 
& Case type 
& Substantive legal category extracted from the document. 
Categories are tort, contract, employment, intellectual property, business and corporate, bankruptcy, and property and real estate. \\

Basics 
& Document type 
& Type of legal filing. 
Categories are pleadings, motions and briefs, affidavits and declarations, court orders and judgments, discovery, exhibits and appendices, settlement and post-judgment filings, and administrative or service documents. \\

Basics 
& Source party 
& Party or institution submitting the document. 
Categories are plaintiff-side, defendant-side, court, and other. \\

Basics 
& Plaintiff type 
& Type of plaintiff identified from the caption and document text. 
Categories are one individual, multiple individuals, corporate, and government. \\

Basics 
& Defendant type 
& Type of defendant identified from the caption and document text. 
Categories are one individual, multiple individuals, corporate, and government. \\

Basics 
& Monetary award requested 
& Dollar amount expressly requested in the document, if stated. \\

Basics 
& Insurance involvement 
& Indicator for whether an insurer or liability policy is explicitly identified in connection with defense, indemnity, coverage, or duty to defend. \\

Basics 
& Independent witness statements 
& Indicator for whether the document references neutral third-party witness statements. \\

Basics 
& Photographic or video evidence 
& Indicator for whether the document references photographic, video, surveillance, dashcam, bodycam, or similar media evidence. \\

Basics 
& Liability admission 
& Indicator for whether the document contains a formal admission of fault by a defendant or authorized representative. \\

Basics 
& Hearing occurred 
& Indicator for whether the document states that a court hearing has already occurred. \\

Basics 
& Motion to dismiss outcome 
& Status or ruling on a motion to dismiss. 
Categories are granted, denied, granted in part, denied in part, withdrawn, pending, and not filed. \\

Basics 
& Summary judgment outcome 
& Status or ruling on a summary judgment motion. 
Categories are granted, denied, granted in part, denied in part, withdrawn, pending, and not filed. \\

Basics 
& Number of plaintiffs 
& Count of named plaintiffs in the caption at the time of filing. \\

Basics 
& Number of defendants 
& Count of named defendants in the caption at the time of filing. \\

Basics 
& Case status 
& Procedural status evidenced by the document. 
Categories are active, stayed, pending settlement, dismissed with prejudice, dismissed without prejudice, judgment entered, settled, and appealed. \\

Basics 
& Document length 
& Length of the OCR-extracted document text. \\

Basics 
& Number of citations 
& Number of legal citations extracted from the document. \\

Basics 
& Relief sought 
& Primary form of relief requested in the document. 
Categories are dismissal, summary judgment, injunctive relief, discovery-related relief, sanctions or fees, arbitration award relief, declaratory or equitable relief, other, and missing or unclear. \\

Basics 
& Number of plaintiff law firms 
& Number of law firms identified on the plaintiff side. \\

Basics 
& Number of defendant law firms 
& Number of law firms identified on the defendant side. \\

Basics 
& Duration 
& Elapsed time between the start of the case and the filing date of the current document. \\

Basics 
& Average wait 
& Average waiting time between filings observed up to the current document. \\

Basics 
& Maximum wait 
& Maximum waiting time between filings observed up to the current document. \\

Basics 
& Document number 
& Ordinal position of the document within the case sequence. \\

Facts 
& Document professionalism 
& LLM-coded assessment of formatting, clarity, and legal reasoning. 
Categories are high, typical, and low. \\

Facts 
& Factual specificity 
& LLM-coded assessment of the specificity of factual allegations. 
Categories are high, typical, and low. \\

Facts 
& Boilerplate fraction 
& Estimated fraction of the document consisting of generic legal standards or copied rule text rather than case-specific argument. \\

Facts 
& Document specificity 
& LLM-derived measure of how case-specific the filing is. \\

Facts 
& Full text embedding dimensions 0--7 
& Eight-dimensional UMAP-compressed representation of the full document text. 
The original full-text representation is a 384-dimensional Sentence-BERT embedding computed by chunking long documents and aggregating chunk embeddings. \\

Facts 
& Summary text embedding dimensions 0--7 
& Eight-dimensional UMAP-compressed representation of the LLM-generated document summary. 
The original summary representation is a 384-dimensional Sentence-BERT embedding. \\

Facts 
& Past win rate (10/50/1,000 NN per full-text/summary embedding) 
& Average plaintiff win rate among the 10/50/1,000 nearest previously resolved cases in full-text/summary embedding space. \\

Facts 
& Minimum case distance (10/50/1,000 NN per full-text/summary embedding)
& Minimum cosine distance between the current document and its 10/50/1,000 nearest previously resolved cases in full-text/summary embedding space. \\

Facts 
& Mean case distance (10/50/1,000 NN per full-text/summary embedding)
& Mean cosine distance between the current document and its 10/50/1,000 nearest previously resolved cases in full-text/summary embedding space. \\

Party information 
& Defendant/Plaintiff law firm experience (1,000 NN per full-text/summary embedding) 
& Average prior experience of defendant-side/plaintiff-side law firms among similar previously resolved cases identified in full-text/summary embedding space. \\

Party information 
& Defendant/Plaintiff law firm win rate (1,000 NN per full-text/summary embedding) 
& Historical defendant-side/plaintiff-side law firm win rate among similar previously resolved cases identified in full-text/summary embedding space. \\

Party information 
& Judge win rate (1,000 NN per full-text/summary embedding) 
& Historical plaintiff win rate in similar previously resolved cases associated with the handling judge. \\

\end{longtable}

\section{Predictive model specification}
\label{app:model}

We model lawsuit outcomes as a three-class classification problem with labels corresponding to plaintiff loss, settlement, and plaintiff win.
The unit of prediction is a document-level observation, indexed by case ID, document number, and filing date.
For each observation, the model receives the feature vector described in \ref{app:features} and outputs predicted probabilities for the three outcome classes.

We use gradient-boosted decision trees implemented with XGBoost as the main predictive model.
The model is trained with the multiclass soft-probability objective and multinomial log-loss as the internal evaluation metric.
We use 500 boosting rounds, histogram-based tree construction, row subsampling of 0.8, a minimum child weight of 5, and feature subsampling selected by cross-validation.
Hyperparameters are selected by grid search over tree depth, learning rate, and column subsampling.
Specifically, we evaluate maximum tree depth in \(\{2,3\}\), learning rate in \(\{0.03,0.05\}\), and column subsampling in \(\{0.6,0.8\}\).

Hyperparameter selection is performed using three-fold grouped cross-validation on the training set.
Folds are grouped by case ID, ensuring that documents from the same case never appear in different cross-validation folds.
The scoring metric for model selection is macro-averaged F1, which weights the three outcome classes equally.
The best-performing specification uses maximum depth 3, learning rate 0.05, and column subsampling 0.6, achieving a mean cross-validated macro-F1 of 0.595.

The temporal train--test split is performed at the case level.
After categorical encoding and feature selection, the all-feature model uses 98 encoded features.
The training set contains 691{,}746 document-level observations from 82{,}176 cases, and the held-out test set contains 143{,}444 observations from 20{,}545 cases.

On the held-out test set, the all-feature XGBoost model achieves an overall accuracy of 0.61, macro-F1 of 0.58, and weighted-F1 of 0.60.
Class-specific F1 scores are 0.53 for plaintiff loss, 0.68 for settlement, and 0.53 for plaintiff win.
Settlement is identified with relatively high recall (0.84), while plaintiff wins and losses exhibit higher precision but lower recall.

For comparison, we also train a model using only the basic structured feature set.
This model uses the same temporal split, model class, and cross-validation procedure, but excludes fact-based text as well as party-level information.
The basic-feature model achieves an overall accuracy of 0.58, macro-F1 of 0.55, and weighted-F1 of 0.56 on the same held-out test set.
Thus, richer fact- and party-based features improve predictive performance relative to structured metadata alone, while the difference remains moderate in aggregate metrics.
This comparison motivates the case-level analyses in the main text, where we examine how the marginal value of richer features varies with case complexity.

\section{Robustness and Ablation Analyses}
\label{app:robustness}

To assess the robustness of the main findings, we conduct two sets of ablation analyses. 
The first examines predictive performance across alternative model classes and train--test designs. 
The second examines whether the complexity-based results depend on the specific predictive model or uncertainty measure used in the main analysis.

\subsection{Prediction ablation}

Our baseline specification uses an XGBoost classifier trained on the full feature set and evaluated using the temporal train--test split described in Section~\ref{sec:data}. 
Table~\ref{tab:prediction_ablation} compares this benchmark against several alternative specifications. 
First, we replace the gradient-boosted model with two alternative classifiers: a Ridge-regularized multinomial regression and a random forest. 
Second, we perturb the train--test design itself. 
We repeat the temporal split using only a random half of the available states, and we additionally consider a more demanding geographic holdout design in which the model is trained on one subset of states and evaluated on a disjoint subset of states. 

\begin{table}[!htbp]
\centering
\scriptsize
\setlength{\tabcolsep}{4pt}
\caption{Prediction ablation study}
\label{tab:prediction_ablation}

\begin{tabular*}{\columnwidth}{@{\extracolsep{\fill}}
l
l
S[table-format=1.3]
S[table-format=1.3]
S[table-format=1.3]
@{}}
\toprule
Configuration                  & Model          & {Accuracy} & {$F1_{\text{micro}}$} & {$F1_{\text{macro}}$} \\
\midrule
Temporal split, full data      & XGBoost        & 0.611 & 0.611 & 0.574 \\
Temporal split, full data      & Ridge          & 0.583 & 0.583 & 0.559 \\
Temporal split, full data      & Random forest  & 0.627 & 0.627 & 0.597 \\
Temporal split, half the states& XGBoost        & 0.628 & 0.628 & 0.562 \\
Train states A, test states B  & XGBoost        & 0.544 & 0.544 & 0.536 \\
\bottomrule
\end{tabular*}
\end{table}

Across specifications, predictive performance remains consistently above chance. 
The linear Ridge model achieves somewhat lower performance than the nonlinear tree-based models, indicating that part of the predictive signal is linearly recoverable but that nonlinear interactions improve classification accuracy. 
The random forest achieves performance comparable to, and slightly above, the baseline XGBoost specification. 
Restricting the analysis to a random subset of states produces similar results, suggesting that the findings are not driven by one particular geographic composition of the sample. 
Performance declines more noticeably under the cross-state holdout design, as expected under a stronger distribution shift, but remains predictive despite the absence of shared jurisdictions between training and test data. 

Overall, the predictive patterns documented in the main text are robust across alternative model classes and train--test designs.

\subsection{Complexity ablation}

The main analysis defines case complexity using the entropy of the predicted outcome distribution generated by the XGBoost classifier. 
To test whether our results depend on this particular combination of predictive model and uncertainty measure, we simultaneously modify both components of the construction. 
Specifically, we replace the baseline XGBoost classifier with a Ridge classifier and redefine complexity using the alternative uncertainty measure
$
1 - \max_i p_i,
$
where $p_i$ denotes the predicted probability assigned to outcome class $i$. 
This quantity provides a simpler measure of predictive uncertainty: complexity is low when one outcome receives most of the probability mass and high when predicted probabilities are more evenly distributed across outcomes. 

Using this alternative setup, we reproduce the three main complexity-related analyses from the paper. 
First, we compare predictive disagreement and accuracy between the baseline-feature model and the full-feature model as a function of complexity. 
Second, we examine the relationship between case duration and complexity. 
Third, we analyze settlement rates as a function of complexity. 

\begin{figure*}[htb]
    \centering
    \includegraphics[width=\linewidth]{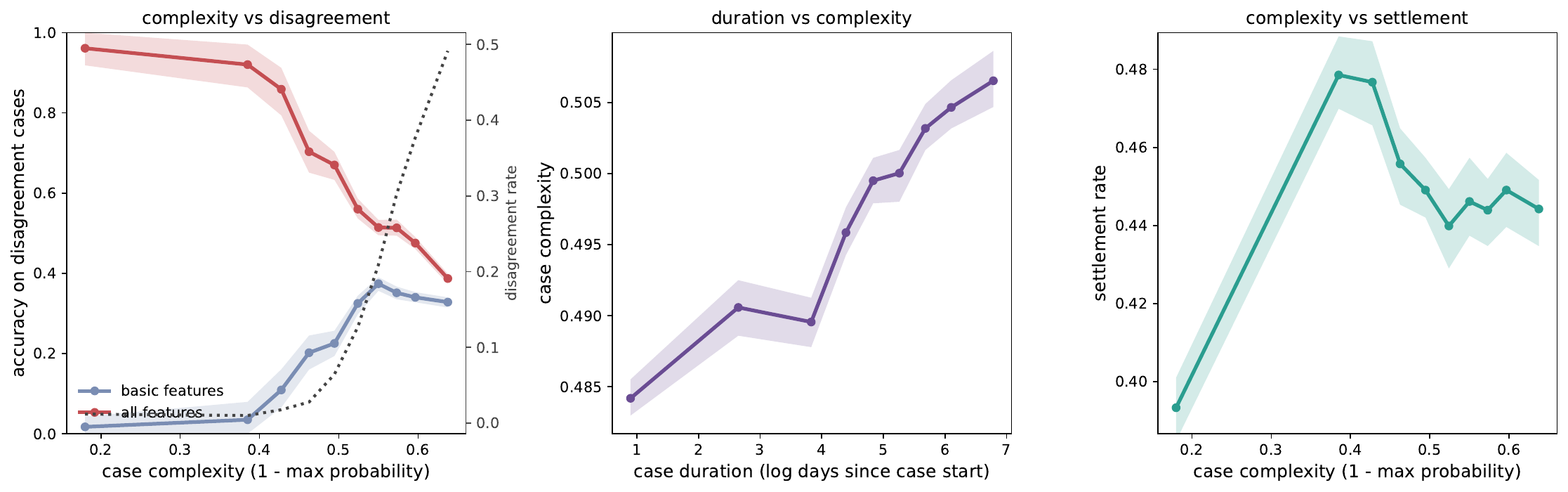}
	\caption{
	Robustness of the complexity analyses under an alternative model and uncertainty measure.
	The figure reproduces the main complexity-related analyses using a Ridge classifier instead of XGBoost and  \(1-\max_i p_i\) instead of entropy as a complexity measure. 
	(Left panel)
	Predictive disagreement and accuracy of the baseline-feature and full-feature models as a function of complexity.
	(Middle panel)
	Average complexity as a function of case duration.
	(Right panel)
	Settlement rate as a function of case complexity.
	The qualitative patterns remain consistent with those reported in the main text.
	}
    \label{fig:complexity_ablation}
\end{figure*}

The qualitative patterns remain stable across all three exercises. 
More complex cases remain associated with greater predictive disagreement, diminishing gains from richer features, longer cases, and systematic variation in settlement behavior. 
These results indicate that the central findings of the paper are not artifacts of the particular classifier or entropy-based uncertainty measure used in the baseline specification.

\end{document}